\documentclass{ws-procs961x669}            

\newcommand{\cH}{\mathcal{H}}
\newcommand{\rL}{\rho_\Lambda}
\newcommand{\CC}{\Lambda}
\newcommand{\xib}{\overline{\xi}}
\newcommand{\bk}{{\bf k}}
\newcommand{\rv}{\rho_{\rm vac}}
\newcommand{\nueff}{\nu_{\rm eff}}

\begin{document}

\title{Vacuum energy and cosmological constant in QFT in curved spacetime}

\author{Joan Solà Peracaula$^{a}$\footnote{Speaker.
 Invited talk  at MG 17, Pescara, Italy, July 7-12, 2024}
 and Cristian Moreno-Pulido$^{b}$}

\address{$^a$Departament de F\'isica Qu\`antica i Astrof\'isica,
and
Institute of Cosmos Sciences,
Universitat de Barcelona,
Av. Diagonal 647, E-08028 Barcelona, Catalonia, Spain\\
$^{b}$ Departament d’Informatica, Matemàtica Aplicada i Estadística,
Universitat de Girona,\\
 17003 Girona, Spain\\
E-mails: sola@fqa.ub.edu, cristian.moreno@udg.edu}
\begin{abstract}

The cosmological constant term (CC), $\CC$, is a pivotal ingredient in the standard model of cosmology or $\CC$CDM, but it is a rigid quantity for the entire cosmic history. This is unnatural and inconsistent. Different theoretical and phenomenological conundrums suggest that the $\CC$CDM necessitates further theoretical underpinning to cope with modern observations. An interesting approach is the framework of the `running vacuum model' (RVM). It endows $\Lambda$  with cosmic dynamics within a  fundamental framework since it is based on QFT. In the RVM, the vacuum energy density (VED) appears as a series of powers of the Hubble function and its derivatives, $\rv (H, \dot{H},...)$.  In the current universe, $\rv$ changes as $\sim H^2$. Higher order effects ${\cal O}(H^4)$, on the other hand, can be responsible for a new mechanism of inflation (RVM-inflation).  On the practical side the RVM can alleviate the cosmological tensions on $\sigma_8$ and $H_0$. An intriguing smoking gun signature of the RVM is that its equation of state can mimic quintessence, as recently observed by DESI, so the vacuum can be the sought-for dynamical DE.  At a deeper theoretical level,  the RVM-renormalized form of the VED  can avoid extreme fine tuning related to the well-known cosmological constant problem.  Overall, the RVM has the capacity  to impinge positively on relevant theoretical and practical aspects of modern cosmology.

\end{abstract}


\bodymatter

\section{Introduction}\label{Sec1:Intro}

The peak point  of the cosmological constant (CC) term,  $\CC$,  in Einstein's equations was its actual measurement at the end of 20th century  using type Ia Supernovae (SnIa) luminosity distances.  The nonvanishing  positive value that was obtained provided for the first time evidence of the accelerated expansion of the Universe\,\cite{SNIa}.  As of its introduction by Einstein in the field equations\,\cite{Einstein1917}, $\CC$ remains a fundamental ingredient of GR. And after more than a century it is still at the core of modern cosmology since it is a key element of the  concordance model of cosmology, $\CC$CDM\,\cite{Peebles1993}. Despite the more general notion of Dark Energy (DE) was introduced much later to `explain' the speeding up of the universe, within the standard model $\CC$ provides the simplest candidate for DE.
Nowadays, very precise measurements have been performed\,\cite{PlanckCollab} on $\CC$ or, equivalently, on the value of the associated parameter $\Omega^0_{\rm vac}=\rho^0_{\rm vac}/\rho^0_{c}$. The vacuum energy density (VED) is associated to $\Lambda$ through the relation $\rho^0_{\rm vac}=\CC /(8\pi G_N)$, where $G_N$ is Newton's constant and $\rho^0_{c}=3H_0^2/(8\pi G_N)$ is the critical density in our time, where $H_0$ is the current value of the Hubble parameter. The observational value of $\Omega^0_{\rm vac}$ is found to be around $0.7$.

Despite $\CC$ is taken from granted within the $\CC$CDM, it undercovers a great source of mystery. To start with, its nature and its origin remain undisclosed at the fundamental level. Naive estimations within quantum field theory (QFT) lead to `gargantuan' values that cannot be minimally reconciled with observations. This abhorrent mismatch is usually said to stem from calculations of the zero-point energy (ZPE) associated to matter fields and also from the Higgs potential of the Standard Model (SM) of particle physics. If $m$ is the mass of one of such SM fields, contributions to the ZPE of order $m^4$ are expected\cite{Zeldovich1967}. Since the observational value of the VED lies around $\rho_{\rm vac}\sim10^{-47}$ GeV$^{4}$ in natural units, comparison with  $\sim m^4$ for any particle of the SM, even the lightest ones,  leads to nonsense.  For example, for the electron field we find a difference of 34 orders of magnitude: $\rho_{\rm vac}^{\rm obs}/m_e^4 \sim 10^{-34}$. The problem persists when comparing with the ground state energy of the effective Higgs potential, leadibng to 56 orders of discrepancy: $V_{\rm eff}$, we find $\rv^{\rm obs}/\langle V_{\rm eff}\rangle\sim 10^{-56}$. These two examples, describe in a conspicuous way what is known as the `Cosmological Constant Problem' (CCP)\,\cite{Weinberg89}. Far from being a mere artifact produced by the simplicity of $\CC$ in the cosmological model, the difficulties seem to lie in the very conception of vacuum energy in Quantum Field Theory (QFT) and affects all forms of DE and not just the vacuum energy\cite{PeeblesRatra2003,Padmanabhan2003,Copeland2006,JSPRev2013,JSPRev2015,JSPRev2022}\,\footnote{For an informal, but rather vivid, introduction to the Cosmological Constant Problem, see e.g. \cite{JSPCosmoverse}.}.
Another aspect of this intriguing topic is the fact that, at the present time, $\rv^{\rm obs}$ and the energy density associated to Cold Dark Matter (CDM), $\rho_{\rm CDM}$,  are observed to be of the same order of magnitude, despite the fact that $\rho_{\rm CDM}$ is assumed to decrease with the expansion as $a^{-3}$ ($a$ being the scale factor), whereas $\rv$ maintains constant. This is the so-called cosmic coincidence problem\,\cite{Steinhardt1997}, although not everybody agrees that this is a real problem\,\cite{Ishak2005}.

There are other worrisome problems with the concordance model of more practical nature. For example, there are tensions  with the measurement of the current Hubble parameter $H_0\equiv 100 h$ km/s/Mpc  ($h\simeq 0.7$) and the growth of  LSS structures.  The growth rate is usually tracked by means of the parameters $S_8$ or $\sigma_8$, related to the root mean square fluctuations of the matter density in spheres of size $8h^{-1}$ Mpc. The $\CC$CDM predicts an excess of structure formation at low redshift at a level of  $\sim 2-3\sigma$ with respect to actual measurements.  The $H_0$-tension, on the other hand, is more serious. It involves an acute disagreement  between the value of $H_0$ inferred from  CMB observations (which make use of fiducial $\CC$CDM cosmology) and the corresponding value extracted from the local (distance ladder) measurements.  The two types of measurements of $H_0$ lead to inconsistencies of $\sim 5 \sigma$ c.l.   See e.g. \cite{Perivolaropoulos:2021jda,Abdalla:2022yfr} for reviews on these tensions.   It is still not known if these tensions are the result of systematic errors, but the possibility that new physics may be ultimately responsible for the observed deviations from the $\CC$CDM predictions cannot be excluded at all.

A wide panoply of  strategies have been proposed in the literature to alleviate some of the above tensions, see the above mentioned references. In this work, we review exclusively the running vacuum model (RVM) approach, which provides a fundamental framework to tackle fundamental cosmological problems as well as  the mentioned tensions with tested efficiency, see \cite{RVMPheno1}  and \cite{RVMPheno2,BDRVM}.  Let us also mention a RVM-inspired model ($w$XCDM) \cite{wXCDM} which recently provided a very robust fit to the data, being also consistent with quintessence-like behavior near our time, as reported by DESI\cite{DESI:2024mwx}. The model $w$XCDM is inspired in the old $\CC$XCDM framework\,\cite{LXCDM}, a composite DE system in which  the RVM is entangled with an extra $X$ component (behaving as `phantom matter') with which it can exchange energy. The idea of `composite running vacuum' appears  fruitful and highly efficient to account for the observations and to cut down the tensions, see \,\cite{SolaPeracaula:2024iil} and references therein.

As said, in this work  we review recent advances in the arduous task of understanding the CCP and the role played by the vacuum in cosmology. Our framework is QFT. We shall, however, not consider quantum gravity here, but QFT in cosmological spacetime,  specifically  Friedmann-Lema\^\i tre-Robertson-Walker (FLRW) spacetime with flat three-dimensional hypersurfaces. The main results have been presented in\,\cite{CristianJoan2020,CristianJoan2022a, CristianJoan2022b,CristianJoanSamira2023}, see also \cite{JSPRev2022} for a comprehensive review.  These studies show that the VED is dynamical in QFT since it depends on the Hubble rate and its derivatives,  $\rv=\rv (H, \dot{H},...)$-- denoted hereafter simply as  $\rv (H$). These studies  also show that contrary to naive expectations there are no Zeldovich type terms of the form $\sim m^4$ like those mentioned at the beginning.  Hence we can avoid extreme fine tuning, in contradistinction to the standard folklore on these matters.   The RVM thus leads to an effective form of the $\CC$CDM in which the physical value of $\CC$  `runs' smoothly with the cosmic expansion thanks to the quantum matter effects.  Such a running, in fact, can be conceived as a renormalization group running, see\,\cite{JSPRev2013} and references therein. As a result,  the quantum vacuum  does not remain rigid throughout the cosmic history. To put it in a nutshell: there is no such thing as a `cosmological constant' in the QFT context.  Interestingly, the running nature of the vacuum has also been accounted for in the context of low-energy effective
string theory\,\cite{ReviewNickJoan2021,PhantomVacuum2021, BasMavSol,NickPhiloTrans,Gomez-Valent:2023hov,Dorlis2024} \footnote{ See also the forthcoming comprehensive review \cite{NickJoan_PR}.}.  In all these formulations, the physical $\CC$ appears as a mildly running quantity since it originates from quantum effects. Nonetheless the consequences on the  $\CC$CDM behavior with a running $\CC$ are not negligible; and in fact they can be essential to cure the cosmological tensions. Here, however, we shall exclusively focus on the QFT approach. For a more comprehensive review encompassing as well the low energy stringy effects on the vacuum dynamics, see \cite{NickJoan_PR}.

Despite many curious attempts existing in the literature, to discuss about the CCP in flat spacetime makes no sense at all. Thus, in what follows we study the RVM renormalization of the energy-momentum tensor (EMT) in cosmological FLRW spacetime. We demonstrate that the values of the VED at different scales are related by a smooth function of the Hubble rate, $H$, and this function does not contain quartic powers $\sim m^4$ of the masses, a fact that drastically eliminates the need for fine tuning. We also discuss the RVM mechanism for inflation  based on the higher powers of $H$. In addition, we show the consequences for the equation of state of the vacuum in an expanding universe, which no longer remains near -1 but can display quintessence-like behavior owing to the quantum corrections.

\section{EMT for a non-minimally coupled scalar field}\label{sec:EMT}

As previously noticed, we wish to address the renormalization of the VED in QFT in curved spacetime. We start with the EMT of the classical field theory.
For the sake of a simpler presentation of the basic ideas, we will assume here that there is only one (matter) field in the form of a real scalar field\footnote{One can generalize the discussion for an arbitrary number of quantized scalar fields and even an arbitrary number of quantized fermions fields, see\,\cite{CristianJoanSamira2023} for details.}  $\phi$. Let us  denote by $T_{\mu \nu}^{\phi}$ the piece of the EMT associated to it Thus $T^{\rm tot}_{\mu\nu}=T_{\mu \nu}^\CC +T_{\mu \nu}^{\phi}$. We neglect other incoherent matter contributions such as dust and radiation because they can be added without altering the QFT considerations developed here.

A non-minimal coupling between the scalar field and gravity is assumed, without any classical potential for $\phi$ . This means that we will restrict ourselves to address the zero-point energy (ZPE) of that field since the ZPE is a common feature of all fields irrespective of their spins and hence is a generic component of the VED.  The part of the action associated to $\phi$ is
\begin{equation}\label{eq:Sphi}
  S[\phi]=-\int d^4x \sqrt{-g}\left(\frac{1}{2}g^{\mu \nu}\partial_{\nu} \phi \partial_{\mu} \phi+\frac{1}{2}(m^2+\xi R)\phi^2 \right)\,,
\end{equation}
where $\xi$ is the non-minimal coupling between $\phi$ and gravity.  In the special case $\xi=1/6$, the massless ($m=0$)  action has conformal symmetry, i.e. symmetric under simultaneous rescalings of the $g_{\mu\nu}$ and $\phi$ with a local function $\alpha(x)$: $g_{\mu\nu}\to e^{2\alpha(x)}g_{\mu\nu}$ and  $\phi\to e^{-\alpha(x)}\phi$. However, we will keep $\xi$ general.

The Klein-Gordon (KG) equation for $\phi$  derived from the above action is
\begin{equation}
(\Box-m^2-\xi R)\phi=0.\label{KGequation}
\end{equation}
Here $\Box\phi=g^{\mu\nu}\nabla_\mu\nabla_\nu\phi=(-g)^{-1/2}\partial_\mu\left(\sqrt{-g}\, g^{\mu\nu}\partial_\nu\phi\right)$. On the other hand, the EMT takes on the form
\begin{equation}
\begin{split}
T_{\mu \nu}(\phi)=&-\frac{2}{\sqrt{-g}}\frac{\delta S_\phi}{\delta g^{\mu\nu}}= (1-2\xi) \partial_\mu \phi \partial_\nu\phi+\left(2\xi-\frac{1}{2} \right)g_{\mu \nu}\partial^\sigma \phi \partial_\sigma\phi\\
& -2\xi \phi \nabla_\mu \nabla_\nu \phi+2\xi g_{\mu \nu }\phi \Box \phi +\xi G_{\mu \nu}\phi^2-\frac{1}{2}m^2 g_{\mu \nu} \phi^2.
\end{split} \label{EMTScalarField}
\end{equation}
In this work we consider a spatially flat FLRW metric in the conformal frame. Denoting by $\tau$ the conformal time, we have  $ds^2=a^2(\tau)\eta_{\mu\nu}dx^\mu dx^\nu$, with $\eta_{\mu\nu}={\rm diag} (-1, +1, +1, +1)$ the Minkowski  metric.
Differentiation with respect the conformal time is denoted by $\left(\right)^\prime\equiv d\left(\right)/d\tau$. It is convenient to define $\cH(\tau)\equiv a^\prime /a$ and since $dt=a d\tau$, we have $\mathcal{H}(\tau)=a  H(t)$, where  $H(t)=\dot{a}/a$  is the usual Hubble rate for the cosmic time.

 \section{From classical to quantum field theory}\label{sec:AdiabaticVacuum}

Next we move to the quantum theory. The ZPE associated with the quantum fluctuations of the field $\phi$ is, of course, UV divergent and requires renormalization in some appropriate scheme. We will \textit{not} use Minimal Subtraction (MS)\cite{Collins84} because it is too artificial to deal with the CCP and involves arbitrary additive constants.  We will use instead a modified form of the method of adiabatic regularization and renormalization\,\cite{BirrellDavies82,ParkerToms09}. The prescription to define the renormalized VED with the new method was implemented in \cite{CristianJoan2020,CristianJoan2022a}.

At the quantum level the EMT decomposes itself as $\langle T_{\mu \nu}^\phi \rangle=\langle T_{\mu \nu}^{\phi_b} \rangle+\langle T_{\mu \nu}^{\delta_\phi}\rangle$, where
$\langle T_{\mu \nu}^{\phi_{b}} \rangle =T_{\mu \nu}^{\phi_{b}} $
is the  contribution  from the background part and  $\langle T_{\mu \nu}^{\delta\phi}\rangle$ is related with the quantum fluctuations of $\phi$. In particular,  $\langle T_{00}^{\delta\phi}\rangle$ is associated with the ZPE of the scalar field in the FLRW background. Thus, the total vacuum contribution to the EMT reads
\begin{equation}
\langle T_{\mu \nu}^{\rm vac} \rangle= T_{\mu \nu}^\CC +\langle T_{\mu \nu}^{\delta \phi}\rangle=-\rL g_{\mu \nu}+\langle T_{\mu \nu}^{\delta \phi}\rangle\,.\label{EMTvacuum}
\end{equation}
Notice that $\rL$ is nothing but the bare term in the Einstein-Hilbert (EH) action.  The vacuum EMT defined above receives contributions from such bare term  and from the quantum fluctuations, both formally divergent. Therefore, to construct the physical VED, $\rv$ (and the physical cosmological term associated with it)  we  need to perform the renormalization of the theory. So at this point the parameters of the EH action and the vacuum EMT are just formal quantities.    The physical VED will emerge only  after providing a renormalized version of this equation which is free from arbitrary constants.

Since the EMT\,\eqref{EMTScalarField} is a combination of different field bilinears, we need to study first the fluctuations $\delta \phi$. The KG equation (\ref{KGequation}) is satisfied independently by the classical field and the quantum fluctuations. The decomposition in Fourier frequency modes $h_k(\tau)$, then, can be cast as
\begin{equation}
\delta \phi(\tau,{\bf x})=\frac{1}{(2\pi)^{3/2}a}\int d^3{k} \left[ A_\bk e^{i{\bf k\cdot x}} h_k(\tau)+A_\bk^\dagger e^{-i{\bf k\cdot x}} h_k^*(\tau) \right]\,. \label{FourierModes}
\end{equation}
$A_\bk$ and  $A_\bk^\dagger $ are the (time-independent) annihilation and creation operators, with usual commutation relations. The KG then translates into
\begin{equation}
h_k^{\prime \prime}+ \Omega_k^2 h_k=0\,, \ \ \ \ \ \ \ \ \ \ \Omega_k^2(\tau) \equiv \omega_k^2(m)+a^2\, (\xi-1/6)R\,, \label{KGModes}
\end{equation}
with  $\omega_k^2(m)\equiv k^2+a^2 m^2$.
As we can see, $h_k$ depends on the modulus $k\equiv|\bk|$ of the momentum. Since $\Omega_k(\tau)$ is a nontrivial function of the conformal time, there is no closed form of the solution to (\ref{KGModes}). Thus, we will generate an approximate solution from  a recursive method based on the phase integral ansatz
\begin{equation}
h_k(\tau)\sim\frac{1}{\sqrt{W_k(\tau)}}\exp\left(i\int^\tau W_k(\tilde{\tau})d\tilde{\tau} \right)\,. \label{WKBSolution}
\end{equation}
Therefore, the function $W_k(\tau)$ satisfies the following differential equation:
\begin{equation}
W_k^2=\Omega_k^2 -\frac{1}{2}\frac{W_k^{\prime \prime}}{W_k}+\frac{3}{4}\left( \frac{W_k^\prime}{W_k}\right)^2\,. \label{WKBIteration}
\end{equation}
This non-linear differential equation can be solved using the WKB approximation. The solution is valid for large $k$ (i.e. for short wave lengths) and the function  $\Omega_k$ is slowly varying for weak fields. The notion of vacuum we work with it is called the adiabatic vacuum -- see \cite{BirrellDavies82,ParkerToms09} and references therein. It is defined as the quantum state annihilated by all the operators $A_k$ of the Fourier expansion of the scalar field.  The physical implications of the theory must be sought in terms of physical observables. However, the physical interpretation of the modes (\ref{KGModes}) with frequencies depending on time is tricky; that is why the dynamics must be analyzed in terms of field observables rather than in particle language. Thus, the physics resides in the  renormalized EMT in the FLRW spacetime. But first we need to regularize it.

\section{Adiabatic Regularization of the EMT}\label{sec:AREMT}

The adiabatic (slowly varying) effective frequency mode $W_k$ defined in \eqref{WKBSolution}, can be computed by means of an  asymptotic series solution of  Eq.\,(\ref{WKBIteration}). This series leads to the  adiabatic regularization procedure (ARP), see \cite{BirrellDavies82,ParkerToms09} and references therein. In the work\cite{Bunch1980}, the method was generalized for scalar fields with arbitrary coupling to the scalar curvature, and more recently it has been further modified for the coupling renormalization in curved backgrounds\,\cite{Ferreiro2019} and  even extended to  spin one-half fields\,\cite{Landete2014}. However, its application to the renormalization of the VED both for scalar fields and spin one-half fields was first performed in\,\cite{CristianJoan2020,CristianJoan2022a,CristianJoan2022b} and \cite{CristianJoanSamira2023}, respectively,  where it was shown that the adiabatic renormalization of the  VED using an off-shell subtraction procedure reproduces exactly the RVM formulas that already existed (in part) in the literature -- see \cite{JSPRev2013} and references therein\,\footnote{For instance, the approach of\cite{KohriMatsui2017} using ARP lacks of the off-shell subtraction procedure,  a fundamental new ingredient\cite{CristianJoan2020,CristianJoan2022a}, and as a consequence the final results presented there still lead to the unwanted $\sim m^4$ contributions triggering the need for the extreme fine tuning characteristic of the CCP. By the same token, the RVM formulas cannot be reached either in that approach.}.  This has opened new vistas for a fresh new look to the cosmological constant problem\cite{JSPRev2022,JSPCosmoverse}.

The aforementioned expansion solution of the nonlinear (WKB-type) equation \eqref{WKBIteration} is  organized in what is known as adiabatic orders\cite{BirrellDavies82,ParkerToms09}. First, the quantites considered of adiabatic order 0 are: $k^2$ and $a$. Of adiabatic order 1 are: $a^\prime$ and $\cH$. Then $a^{\prime \prime},a^{\prime 2},\cH^\prime$ and $\cH^2$ are quantities of adiabatic order 2. We can sum up by saying that each extra derivative in conformal (or cosmic) time increases the adiabatic order one unit. As a consequence, the  ``effective frequency'' $W_k$ can be written as an asymptotic expansion:
\begin{equation}\label{WKB}
W_k=\omega_k^{(0)}+\omega_k^{(2)}+\omega_k^{(4)}+\dots,
\end{equation}
where each $\omega_k^{(j)}$ is an adiabatic correction of order $j$.  This leads to an expansion of the mode function $h_k$ in even order  adiabatic terms. This is justified by arguments of general covariance, since only terms of even adiabatic order (an even number of time derivatives) are allowed in the field equations.

\subsection{Introducing the renormalization parameter}\label{sect:RelatingScales}

We start by defining the 0\textit{th} order terms
\begin{equation}\label{omegak0}
\omega_k^{(0)} \equiv\omega_k= \sqrt{k^2+a^2 M^2} .
\end{equation}
In this approach the WKB expansion is performed off-shell, at an arbitrary mass scale $M$ replacing the scalar field mass $m$ in \eqref{omegak0}. In consequence, ARP can be formulated in such a way that we can relate the adiabatically renormalized theory at two different scales. If $M = m$  we obtain the renormalized theory on-shell. In the computation of the EMT, the parameter $\Delta^2\equiv m^2-M^2$ will appear in the correction terms and it has to be considered of adiabatic order 2 since it appears in the WKB expansion together with other terms of the same adiabatic order\,\cite{Ferreiro2019}. If $\Delta = 0$, then $M = m$ and corresponds to the usual on-shell ARP\cite{BirrellDavies82,ParkerToms09}.
With the help of this formalism we explore the evolution of the VED throughout the cosmic history as it was done for the first time in \,\cite{CristianJoan2020}. The crucial point is to define the off-shell subtraction of the EMT at the scale $M$ (see Sec.\ref{sec:RenormZPE})  and then (and no less important) to properly identify its value with a cosmic observable at the corresponding epoch  (in this case the Hubble rate $H$), but only at the end of the calculations. In this way the scale $M$ can be interpreted as a renormalization group scale. For the sake of  simplicity, we model here this procedure only  in terms of real scalar fields as in\,\cite{CristianJoan2020,CristianJoan2022a}. For a generalization to fermion fields, see\,\cite{CristianJoanSamira2023}.
\subsection{Adiabatically regularized ZPE}\label{eq:RegZPE}

Our starting point is the initial solution $W_k\approx \omega_k^{(0)}$ indicated in Eq. \eqref{omegak0}.  For  $a=1$, it corresponds to the standard Minkowski space modes. Since $a=a(\tau)$ we have to find a better approximation.  We use the initial solution $\omega_k^{(0)}$ in  \eqref{WKBIteration} and expand the RHS in powers of  $\omega_k^{-1}$ (hence large momenta $k$), then collect the new terms up to adiabatic order 2 to find $\omega_k^{(2)}$. We repeat the process with $W_k\approx \omega_k^{(0)}+\omega_k^{(2)}$ on the RHS of the same equation, expand again  in  $\omega_k^{-1}$, and collect contributions of adiabatic order 4, 6, ... Each step is increasingly more demanding in the number of terms involved and in practice we do not go beyond 4th or  6th order (the calculation at 6th order is already rather cumbersome but still feasible\,\cite{CristianJoan2022a,CristianJoanSamira2023}). In any case, recall that the series is asymptotic and hence should not be continued indefinitely, even if technically possible.  After a few steps, the expansion emerges organized in powers of  $\omega_k^{-1}\sim 1/k$ (i.e. a short wavelength expansion). The UV divergent terms of the ARP are precisely the first lowest powers of $\omega_k^{-1}$, which are present in the first adiabatic orders of the expansion. Higher adiabatic orders come later in the iteration, and  represent finite contributions since they decay quickly at large $k$ and the associated integrals are manifestly convergent. In our case, the divergent terms of the EMT are present up to 4th  adiabatic order (in $n=4$ spacetime dimensions). This means  that we have to compute all the terms up to this order at least. Upon appropriate subtraction at the scale $M$, we will obtain a finite expression for the EMT and then compute the vacuum energy density.
After performing the adiabatic expansion on the different field bilinears, we end up with the unrenormalized ZPE:
\begin{equation}
\begin{split}
\langle T_{00}^{\delta \phi} \rangle & =\frac{1}{8\pi^2 a^2}\int dk k^2 \left[ 2\omega_k+\frac{a^4M^4 \cH^2}{4\omega_k^5}-\frac{a^4 M^4}{16 \omega_k^7}(2\cH^{\prime\prime}\cH-\cH^{\prime 2}+8 \cH^\prime \cH^2+4\cH^4)\right.\\
&+\frac{7a^6 M^6}{8 \omega_k^9}(\cH^\prime \cH^2+2\cH^4) -\frac{105 a^8 M^8 \cH^4}{64 \omega_k^{11}}\\
&+\overline{\xi}\left(-\frac{6\cH^2}{\omega_k}-\frac{6 a^2 M^2\cH^2}{\omega_k^3}+\frac{a^2 M^2}{2\omega_k^5}(6\cH^{\prime \prime}\cH-3\cH^{\prime 2}+12\cH^\prime \cH^2)\right. \\
& \left. -\frac{a^4 M^4}{8\omega_k^7}(120 \cH^\prime \cH^2 +210 \cH^4)+\frac{105a^6 M^6 \cH^4}{4\omega_k^9}\right)\\
&+\left. \xib^2\left(-\frac{1}{4\omega_k^3}(72\cH^{\prime\prime}\cH-36\cH^{\prime 2}-108\cH^4)+\frac{54a^2M^2}{\omega_k^5}(\cH^\prime \cH^2+\cH^4) \right)
\right]\\
&+\frac{1}{8\pi^2 a^2} \int dk k^2 \left[  \frac{a^2\Delta^2}{\omega_k} -\frac{a^4 \Delta^4}{4\omega_k^3}+\frac{a^4 \cH^2 M^2 \Delta^2}{2\omega_k^5}-\frac{5}{8}\frac{a^6\cH^2 M^4\Delta^2}{\omega_k^7} \right.\\
& \left. + \xib \left(-\frac{3a^2\Delta^2 \cH^2}{\omega_k^3}+\frac{9a^4 M^2 \Delta^2 \cH^2}{\omega_k^5}\right)\right]+\dots, \label{EMTFluctuations}
\end{split}
\end{equation}
where we have defined $\xib\equiv \xi-1/6$ and we have integrated $ \int\frac{d^3k}{(2\pi)^3}(...)$ over solid angles and expressed the final integration in terms of $k=|\bk|$. Let us note a couple of things. First, there are terms in the above result which are  manifestly UV-divergent. Second, we note the presence of the $\Delta$-dependent terms in the last two rows, which contribute at 2{\it th }($\Delta^2$) and 4{\it th } ($\Delta^4$) adiabatic order.

\section{Renormalization of the ZPE in curved spacetime}\label{sec:RenormZPE}

The ZPE part of the EMT, as given by  Eq.\,\eqref{EMTFluctuations} can be split into two parts as follows:
\begin{equation}
\langle T_{00}^{\delta \phi}\rangle (M)= \langle T_{00}^{\delta \phi}\rangle_{\rm Div}(M)+\langle T_{00}^{\delta \phi}\rangle_{\rm Non-Div}(M), \label{DecompositionEMT}
\end{equation}
The non-divergent part of (\ref{DecompositionEMT}) involves the integrals with powers of  $1/\omega_k$ higher than $3$ which are perfectly finite. On the other hand, the divergent part is
\begin{equation}
\begin{split}
&\langle T_{00}^{\delta \phi}\rangle_{\rm Div}(M)=\frac{1}{8\pi^2 a^2}\int dk k^2 \Bigg[ 2\omega_k +\frac{a^2 \Delta^2}{\omega_k}-\frac{a^4\Delta^4}{4\omega_k^3}\\
& -6\xib \cH^2\left(\frac{1}{\omega_k}+\frac{a^2 M^2}{\omega_k^3}+\frac{a^2 \Delta^2}{2\omega_k^3}\right) -9\frac{\xib^2}{\omega_k^3}(2\cH^{\prime\prime}\cH-\cH^{\prime 2}-3\cH^4) \Bigg],
\end{split} \label{DivergentPart}
\end{equation}
which contains the powers  $1/\omega_k^n$ up to $n=3$, manifestly UV-divergent.

Let us now focus on the divergent part of the ZPE, Eq.\,\eqref{DivergentPart}.  First, we are going to set the arbitrary scale at the mass of the scalar field, that is, $M=m$,  and hence $\Delta=0$ (cf. Sec.\ref{sect:RelatingScales}). The divergent part (\ref{DivergentPart}) is reduced in this case to
\begin{equation}
\begin{split}
&\langle T_{00}^{\delta \phi}\rangle_{\rm Div}(m)=\frac{1}{8\pi^2 a^2}\int dk k^2 \Bigg[ 2\omega_k(m) -\xib  6\cH^2\left(\frac{1}{\omega_k(m)}+\frac{a^2m^2}{{\omega_k^3(m)}}\right) \\
& -\xib^2\frac{9}{{\omega_k^3(m)}}(2\cH^{\prime\prime}\cH-\cH^{\prime 2}-3\cH^4) \Bigg]\,.
\end{split} \label{DivergentPartClassic}
\end{equation}
Right next, in order to renormalize the ZPE and the EMT, we will subtract the terms that appear up to 4th adiabatic order at the arbitrary mass scale $M$.  This procedure suffices to cancel the divergent terms through the ARP and avoids leaving arbitrary constants in the subtraction procedure (as in the MS scheme\cite{Collins84}).

\subsection{Off-shell renormalized ZPE}\label{sec:ZPEoffshell}

Our specific proposal for the renormalized  ZPE in curved spacetime at the scale $M$ is the subtracted form\cite{CristianJoan2020,CristianJoan2022a}:
\begin{eqnarray}\label{EMTRenormalized}
\langle T_{00}^{\delta \phi}\rangle_{\rm Ren}(M)&=&\langle T_{00}^{\delta \phi}\rangle(m)-\langle T_{00}^{\delta \phi}\rangle^{(0-4)}(M)\nonumber\\
&=&\langle T_{00}^{\delta \phi}\rangle_{\rm Div}(m)-\langle T_{00}^{\delta \phi}\rangle_{\rm Div}(M)-\xib\frac{3\Delta^2 \cH^2}{8\pi^2}+\dots
\end{eqnarray}
Here $(0-4)$ means that the expansion is up to fourth adiabatic order and the dots in (\ref{EMTRenormalized}) denote finite terms of higher adiabatic order that come from the finite part of the EMT.  Using now Eq.\,\eqref{DivergentPartClassic}, we arrive at the expression
\begin{equation}
\begin{split}
&\langle T_{00}^{\delta \phi}\rangle_{\rm Ren}(M)=\frac{1}{8\pi^2 a^2}\int dk k^2 \left[ 2 \omega_k (m)-\frac{a^2 \Delta^2}{\omega_k (M)}+\frac{a^4 \Delta^4}{{4\omega^3_k (M)}}-2 \omega_k (M)\right]\\
&+\xib \frac{6\cH^2}{8\pi^2 a^2}\Bigg\{ \int dk k^2 \left[\frac{1}{\omega_k (M)}+\frac{a^2 M^2}{{\omega^3_k (M)}}+\frac{a^2 \Delta^2}{2{\omega^3_k (M)}}-\frac{1}{\omega_k(m)}-\frac{a^2 m^2}{{\omega^3_k (m)}} \right]-\frac{a^2\Delta^2}{2}\Bigg\}\\
&-\xib^2 \frac{9\left(2 \cH^{\prime \prime}\cH-\cH^{\prime 2}-3 \cH^{4}\right)}{8\pi^2 a^2}\int dk k^2 \left[ \frac{1}{{\omega^3_k (m)}}-\frac{1}{{\omega^3_k (M)}}\right]+\dots
\end{split} \label{Renormalized}
\end{equation}
In this equation we have introduced new notation in order to distinguish between the off-shell energy mode $\omega_k(M)\equiv \sqrt{k^2+a^2 M^2}$  (formerly denoted simply as $\omega_k$) and the on-shell one $\omega_k(m)\equiv \sqrt{k^2+a^2 m^2}$.
Despite individual integrals in \eqref{Renormalized} look UV-divergent, the overall result is finite.  After some due amount of algebra and with the help of Mathematica\cite{Mathematica} the final result can be cast simply as follows:
\begin{equation}
\begin{split}
&\langle T_{00}^{\delta \phi}\rangle_{\rm Ren}(M)=\frac{a^2}{128\pi^2 }\left(-M^4+4m^2M^2-3m^4+2m^4 \ln \frac{m^2}{M^2}\right)\\
&-\xib\frac{3 \cH^2 }{16 \pi^2 }\left(m^2-M^2-m^2\ln \frac{m^2}{M^2} \right)+\xib^2 \frac{9\left(2  \cH^{\prime \prime} \cH- \cH^{\prime 2}- 3  \cH^{4}\right)}{16\pi^2 a^2}\ln \frac{m^2}{M^2}+\dots
\end{split} \label{RenormalizedExplicit2}
\end{equation}
This is the off-shell renormalized ZPE at 4th order depending on the arbitrary scale $M$. We note that it vanishes for $M=m$ as expected from Eq.\,(\ref{EMTRenormalized}). However, let us remind the reader that this only happens when computing the on-shell value $\langle T_{\mu\nu}^{\delta \phi}\rangle_{\rm Ren}(m)$ up to adiabatic order $4$ in Eq.\,(\ref{RenormalizedExplicit2}). Nevertheless, $\langle T_{\mu\nu}^{\delta \phi}\rangle_{\rm Ren}(m)$ can be computed up to an arbitrary, but finite, adiabatic order. Beyond 4th order one always obtains subleading and perfectly finite corrections. At low energies (i.e. in the late universe, where $H\ll m$) we expect that these effects become suppressed.  However, at high energies ($H\sim m$), i.e. in the early universe, these effects can be relevant and  be responsible for a genuinely new inflationary mechanism (RVM-inflation) -- cf. Sec. \ref{sect:inf}.

So far, we have obtained a renormalized ZPE in curved spacetime (\ref{RenormalizedExplicit2}) which, despite the fact of being finite, it still involves the quartic powers $\sim m^4$ of the masses, which entail harmful contributions to the vacuum energy. Fortunately, this is not yet the expression for the renormalized VED, which will be free from them!

\section{Renormalized vacuum energy density}\label{sec:RenormalizedVED}

The renormalization in the context of QFT in curved spacetime implies the consideration of the higher derivative (HD) terms in the classical effective action\,\cite{BirrellDavies82}, beyond the usual EH term with a cosmological constant, $\CC$. In particular, in the FLRW background in four dimensions, the only surviving HD term is $H_{\mu \nu}^{(1)}$, as suggested by the geometric structure of \,(\ref{RenormalizedExplicit2}).  $H_{\mu \nu}^{(1)}$ is  obtained by functionally differentiating  $R^2$ with respect the metric\cite{BirrellDavies82,CristianJoan2022a}.  So, the full action contains the EH+HD terms and also the matter part which, in our simplified case, only consists of a scalar field $\phi$ non-minimally coupled to gravity, Eq.\,(\ref{eq:Sphi}). As usual, the modified Einstein's equations are obtained through the variation of the new action with respect to $g_{\mu\nu}$:
\begin{equation}\label{eq:MEEs}
\mathcal{M}_{\rm Pl}^2 (M) G_{\mu \nu}+\rL(M) g_{\mu \nu}+\alpha(M) H_{\mu \nu}^{(1)}=\langle T_{\mu \nu}^{\delta \phi} \rangle_{\rm Ren}(M)\,,
\end{equation}
where the dependence in $M$ indicates we are considering renormalized quantities and
\begin{equation}
\mathcal{M}_{\rm Pl}^2 (M) \equiv \frac{1}{8\pi G_N(M)}
\end{equation}
is simply the reduced Planck mass at the scale $M$.
Our goal is not to calculate the VED but to relate it at different renormalization points, and with this purpose in mind let us subtract Einstein's equations as written in \eqref{eq:MEEs} at two different scales, $M$ and $M_0$. We  find
\begin{equation}
\langle T_{\mu \nu}^{\delta \phi}\rangle_{\rm Ren}(M)- \langle T_{\mu \nu}^{\delta \phi}\rangle_{\rm Ren}(M_0)=\delta\mathcal{M}_{\rm Pl}^2    G_{\mu \nu}+\delta\rL g_{\mu \nu}+\delta \alpha H^{(1)}_{\mu \nu},  	\label{EinsteinDifferentScale}
\end{equation}
with $\delta X(m,M,M_0)\equiv X(M)- X(M_0)$ for the various parameters involved $X=\mathcal{M}_{\rm Pl}^2, \rL, \alpha$.
Since we know the expression of the renormalized EMT within the ARP, namely  Eq.\, (\ref{RenormalizedExplicit2}),  we can obtain the scaling of the couplings in \eqref{EinsteinDifferentScale} between $M$ and $M_0$.  This can be accomplished by taking the value of the components $G_{00}$ and $H_{00}^{(1)}$ (explicit formulas are given in \cite{CristianJoan2022a}) and comparing with (\ref{RenormalizedExplicit2}).  Notice that the first term on the \textit{r.h.s} of (\ref{RenormalizedExplicit2})  is of 0{\it th} adiabatic order and is associated to $\delta\rL (m,M,M_0)$. Similarly with the others. The result is
\begin{equation} \label{SubtractionOfCoupling}
\begin{split}
&\delta\mathcal{M}_{\rm Pl}^2(m,M,M_0)\equiv \mathcal{M}_{\rm Pl}^2 (M)- \mathcal{M}_{\rm Pl}^2 (M_0) = \frac{\xib}{16\pi^2}\left[M^2 - M_0^{2} -m^2\ln \frac{M^{2}}{M_0^2}\right], \\
&\delta  \rL (m,M,M_0) =\frac{1}{128\pi^2}\left(M^4-M_0^{4}-4m^2(M^2-M_0^{2})+2m^4\ln  \frac{M^{2}}{M_0^2}\right),\\
& \delta \alpha (M,M_0)= -\frac{\xib^2}{32\pi^2} \ln \frac{M^2}{M_0^{2}}.\\
\end{split}
\end{equation}

\subsection{Running Vacuum Energy density. Absence of $\sim m^4$ terms.}\label{sec:TotalVED}

At this point we may come back to the definition of VED in our framework, Eq.\eqref{EMTvacuum}.  We consider $\langle T_{\mu\nu}^{{\rm vac}}\rangle=P_{\rm vac}g_{\mu \nu}+\left(P_{\rm vac}+\rv\right)u_\mu u_\nu$ (i.e. we assume a perfect fluid form for the vacuum but no particular EoS for it). Using $u_\mu=(-a,0,0,0)$ and equating the perfect fluid form to Eq.\,\eqref{EMTvacuum}, and then taking the $00$-component of the equality,
we find $ T_{\mu\nu}^{{\rm vac}}=a^2\rv$ (which holds good for any relationship between $P_{\rm vac}$ and $\rv$, i.e. irrespective of the EoS for the vacuum, which we will compute later on!). Therefore,  we arrive at the renormalized VED:
\begin{equation}\label{eq:Totalrhovac}
\rv(M)=\rL(M)+\frac{\langle T_{00}^{\delta \phi}\rangle_{\rm Ren}(M )}{a^2}.
\end{equation}
This equation shows that the renormalized VED at an arbitrary  scale $M$ is not only receiving contributions from the renormalized cosmological term but also from the quantum fluctuations of fields (i.e. from the renormalized ZPE, which we have computed). It displays also the dependence on both the renormalization scale $M$ and the Hubble function $H$. Using \eqref{RenormalizedExplicit2} we find it explicitly:
\begin{equation}
\begin{split}\label{DependenceMH}
\rho_{\rm vac}(M,H)&=\rho_\Lambda (M)+\frac{1}{128\pi^2}\left(-M^4+4m^2M^2-3m^4+2m^4\ln\frac{m^2}{M^2}\right)\\
&+\bar{\xi}\frac{3H^2}{16\pi^2}\left(M^2-m^2+m^2\ln\frac{m^2}{M^2}\right)+\mathcal{O}(H^4)\,.
\end{split}
\end{equation}
$\mathcal{O}(H^4)$ collects different terms with 4 derivatives of time $(H^4,H^2\dot{H},\dots$). The values of $M$ and $H$ are independent, but a selected choice of the renormalization point $M$ near $H$ corresponds to choose the RG scale around the characteristic energy scale of FLRW spacetime at any given moment of the expansion history, and hence it should have plenty of physical significance. We can indeed  derive the ‘low energy’ (late universe) form of the VED  along these lines.
Upon subtracting the renormalized result at two scales, $M$ and $M_0$, and using the above equations, one finds
\begin{equation}
\begin{split}
\rv(M,H)
=&\rv(M_0,H)+\frac{3}{16\pi^2}\xib H^2\left[M^2 - M_0^{2}-m^2\ln \frac{M^{2}}{M_0^2}\right]+\mathcal{O}(H^4)\,.\label{CosmicTimeRenormalizedVE}
\end{split}
\end{equation}
This equation gives  the value of the VED at the scale $M$, related with the value of the VED at another renormalization scale $M_0$, i.e. it  expresses the `running' of the VED. Such running is manifestly  slow in the current universe and for $\xi=1/6$ ($\xib = 0$) there is no running at all (as expected). Finally, we remark particularly the absence of $\sim m^4$ contributions. Ergo there is no need of extreme fine tuning of the VED that could recreate the CCP in the RVM framework.

\section{The physics of the running vacuum in the current universe}\label{sec:RunningConnection}

Our framework is intended for the study  of the `running' or scaling evolution of the vacuum energy density from one scale to another across the cosmic history. Therefore, if the VED $\rv$  is known at some (arbitrary) initial scale $M_0$, our equations predict the value of $\rv$ at another scale $M$. This can be very useful for cosmology. However the  renormalized result (in any renormalization scheme in QFT)  lacks a  physical interpretation until a clear physical meaning is ascribed to the renormalization scale $M$. It is important to note that  despite the fact that the  full effective action of the theory is independent of the renormalization scale $M$,  the vacuum effective action $W_{\rm eff}$\,\cite{BirrellDavies82} and other sectors of the theory can indeed depend on it. Therefore, it is perfectly possible to test the scaling evolution of different sectors of the theory, in particular the vacuum sector.  Of course, here resides the  key to physics, and this is why the renormalization group method is useful at all. An adequate choice of $M$ at the end of the renormalization process can be obtained if one picks  it equal to the value of the expansion rate,  $H$,  at each cosmic epoch under consideration. It corresponds to choose the renormalization group scale around the characteristic energy scale of  FLRW spacetime at any given moment. This  should have physical significance. In fact, such a setting procedure is entirely similar to the standard practice in ordinary gauge theories, where the choice of  the renormalization scale  is  made near the typical energy of the process. In our case,  the `process' is the cosmic expansion itself, framed  in the FLRW background, and hence its characteristic energy scale at all times should be the very value of $H$ at the corresponding epoch of the cosmic expansion.
Such a connection was suggested long ago from semi-qualitative renormalization group arguments --  see  \cite{JSPRev2013,JSPRev2015}  and references therein for a review of the running vacuum model (RVM)\footnote{Recently, the setting  $M=H$ has also been underpinned within the context of cosmological studies based on lattice Quantum Gravity, see\,\cite{Dai:2024vjc}. In a different vein, it is also useful for the  study of a possible variation of the fundamental `constants' of nature ($\CC, G, \alpha_{\rm em}, \alpha_s, \CC_{\rm QCD},...$)\,\cite{Fritzsch:2012qc,MemorialHF2024} with the evolution of the universe. This is recursively a hot subject in the literature.}. But it was only only after the detailed works \,\cite{CristianJoan2020,CristianJoan2022a, CristianJoan2022b,CristianJoanSamira2023} that it was possible to put it on firm QFT ground.

The value itself of the vacuum energy density cannot be computed from renormalization theory in QFT. So our aim is to relate VED values at different points $H$ and $H_0$ of the cosmic evolution. The current value of the VED, $\rv^0\equiv \rv (H_0)$,  corresponds to fixing the numerical value of the scale at $M=H_0$,  being $H_0$  today's value of the Hubble function. From this setup we can predict the value of the VED at another cosmic epoch $M=H$.  Keeping our focus on cosmic epochs $H$ and $H_0$ accessible to our observations (hence $H$ being a past epoch, $H>H_0$), the shift of the VED immediately follows from \eqref{CosmicTimeRenormalizedVE}:
\begin{equation}
\begin{split}
\rho_{\rm vac}(H) &= \rho_{\rm vac}(H_0)+\frac{3\bar{\xi}}{16\pi^2}\left[m^2H^2\left(-1+\ln \frac{m^2}{H^2}\right)-m^2H_0^2\left(-1+\ln \frac{m^2}{H_0^2}\right)\right]+\dots
\end{split}\label{eq:RunningFormula}
\end{equation}
where $\rho_{\rm vac}(H)\equiv\rho_{\rm vac}(M=H,H)$ and similarly $\rho_{\rm vac}(H_0)\equiv\rho_{\rm vac}(H_0,H_0)$,  the dots representing terms of 4th adiabatic order and beyond. Equation \eqref{CosmicTimeRenormalizedVE} can be used , in principle, to explore the value of the VED throughout the cosmological history, specifically for all post-inflationary times. However, for the study of the very early universe (in particular, for the inflationary stage) the higher order contributions can surely play a role. We address  RVM-inflation briefly in the next section.

The previous formula shows that there is in effect a ‘running’ or change of the VED from $H_0$ to $H$. Notice that if $m$ is an ordinary particle mass (e.g. within the standard model of particle physics) the running would be very small. We can make it more transparent by rewriting the above equation as follows. Define the effective running parameter
\begin{equation}
\nu_{\rm eff}(H) = \frac{1}{2\pi}\bar{\xi}\frac{m^2}{m_{\rm Pl}^2}\left(-1+\frac{m^2}{H^2}-\frac{H_0^2}{H^2-H_0^2}\ln \frac{H^2}{H_0^2}\right).
\end{equation}
The running VED formula \eqref{eq:RunningFormula} can now be written in a rather compact form as follows:
\begin{equation}\label{eq:RVMLate}
\rho_{\rm vac}(H) = \rho_{\rm vac}^0 +\frac{3\nu_{\rm eff}(H)}{8\pi G_{\rm N}}(H^2-H_0^2)\,,
\end{equation}
where $\rho_{\rm vac}^0=\rho_{\rm vac}(H_0)$ is identified with today's VED value, $\rho_{\rm vac}^0$, and $G_{\rm N}$ is assumed to be the currently measured value of the gravitational constant. Clearly, $\nu_{\rm eff}(H)\lll 1$ unless $m$ is a particle mass near a typical GUT scale $M_X\sim 10^{16}$ GeV. Only then the ratio  $m/m_{\rm Pl}$ can be small but not extremely small, especially if taking into account the large multiplicity of particles in a typical GUT. The evolution of $\nu_{\rm eff}(H)$ is logarithmic and very small.  Effectively,
\begin{equation}
\nu_{\rm eff}(H)\approx \nu_{\rm eff}(H_0)=\frac{1}{2\pi}\bar{\xi}\frac{m^2}{m_{\rm Pl}^2}\ln \frac{m^2}{H_0^2}\,.
\end{equation}
One can then estimate $\nu_{\rm eff}(H_0)=10^{-5}-10^{-3}$, see\cite{Fossil07}. The VED is thus expected to evolve very little (logarithmically) during the post-inflationary expansion and should remain close to $\nu_{\rm eff}(H_0)$.
From the foregoing, it follows that it can be treated to
a good approximation as a small parameter within the observable universe. Such approximation holds good even if we study the CMB epoch, since the difference between $\nueff (H)$ ans $\nueff (H_0)$ can be estimated to be less than a few percent. In addition, the  BBN physics remains phenomenologically safe \cite{Asimakis2022}. With $\nu_{\rm eff}$ constant, equation\,\eqref{eq:RVMLate} renders the canonical form of the VED for the running vacuum model RVM\,\cite{JSPRev2013}. Phenomenological studies coincide that the running parameter must be in the approximate range  $10^{-4} - 10^{-2}$ depending on the scenarios-- cf. the various analyses \cite{RVMPheno1,RVMPheno2,BDRVM}. For $\nu_{\rm eff}(H)$ in the mentioned range, the impact on the phenomenological studies can be significant and can even help to solve the cosmological tensions\cite{RVMPheno1}.

\section{RVM-inflation: a genuinely new  mechanism}\label{sect:inf}

The importance of the lower order terms in the adiabatic expansion has been  emphasized in the previous section. Now it is time to focus on those terms beyond second adiabatic order ($H^4$, $H^6$,...) which are totally irrelevant near the present time but nonetheless may play a capital role in the early universe, namely during the inflationary period. The key idea is that a large value of $H=$const. from these terms does magnify the magnitude of the VED in the very early universe, whereby producing a strong boost in the expansion rate for a short period. The power structure of the  RVM can indeed provide a mechanism of inflation that relies on the natural capability of the quantum vacuum to be enhanced in the primordial era. This is quite different from ad hoc (inflaton) fields added in the classical action. It is also very different from the Starobinsky mechanism\cite{Starobinsky80} for which $H\neq$const. (seee the pertinent discussion of Ref.\cite{JSPRev2015}).

In the RVM context, the mechanism of inflation requires the VED to depend on an even power of the Hubble function such as $H^4,H^6,...$ or a linear combination of them.   While these powers were used in previous phenomenological approaches\cite{rvmInflationpheno,Sola:2015csa,Yu2020}, now we have a full theoretical justification for their existence\cite{CristianJoan2020,CristianJoan2022a, CristianJoan2022b}.  We call this QFT-based mechanism `RVM-inflation'. It is characterized by a short period where $H\approx$ const. and the VED is entirely dominated by one of these higher powers. Here we single out power $H^6$ since it is the first one which appears explicitly even before setting $M=H$ in the renormalized VED \footnote{Should we implement the setting $M=H$, the first inflationary power would  be $H^4$. A detailed study of this alternative situation will be presented in \cite{Alex}. This power also emerges in stringy formulations of the RVM\cite{ReviewNickJoan2021,PhantomVacuum2021}. In contrast, for Starobinsky inflation \cite{Starobinsky80} none of these powers appears in the VED, only higher derivatives of $H$, which  just vanish for $H=$const. --  see\cite{JSPRev2015}.}.  The specific form of the $H^6$ term driving RVM-inflation is obtained after a lengthy calculation involving the 6th adiabatic order. The part that remains nonvanishing for $H=$const. is relatively simple and reads as follows,\cite{CristianJoan2022a,CristianJoanSamira2023}
\begin{equation}\label{eq:rhovacinflation}
 \rho_{\rm vac}^{\rm inf} \sim \frac{\langle T_{00}^{\delta \phi} \rangle_{\rm ren}}{a^2}= {C_{\rm inf}} H^6\,,
 \end{equation}
 where $C_{\rm inf}$ is a calculable  coefficient which depends on the non-minimal coupling and on the inverse square of the masses of the fields (bosons and fermions) participating in the calculation -- see\,\cite{CristianJoan2022a,CristianJoanSamira2023} for details. The relevant fields contributing to $C_{\rm inf}$ are those that belong to any typical GUT scale ($M_X \sim 10^{16}$ GeV). Solving the cosmological equations in the presence of the term \eqref{eq:rhovacinflation} and neglecting at this point the low energy contributions to the VED (viz. the constant additive term and the $\sim H^2$ power)  yields a simple analytical solution:
%
 $H(\hat{a}) = H_I \left(1+\hat{a}^8\right)^{-1/4}$,
%
in which $H_{\rm I}$ is the inflationary value at the GUT scale.
The  corresponding energy densities of radiation and vacuum read
\begin{equation}\label{eq:Inflationdensities}
\rho_{\rm vac}(\hat{a}) =  \rho_{\rm I} \left(1+\hat{a}^8\right)^{-3/2}\,, \qquad \rho_{\rm r} = \hat{a}^8 \rho_{\rm vac}\left(\hat{a}\right)\,.
\end{equation}
\begin{center}
\begin{figure}[t]\label{figure:Inflation}
\begin{center}
\includegraphics[scale=0.44]{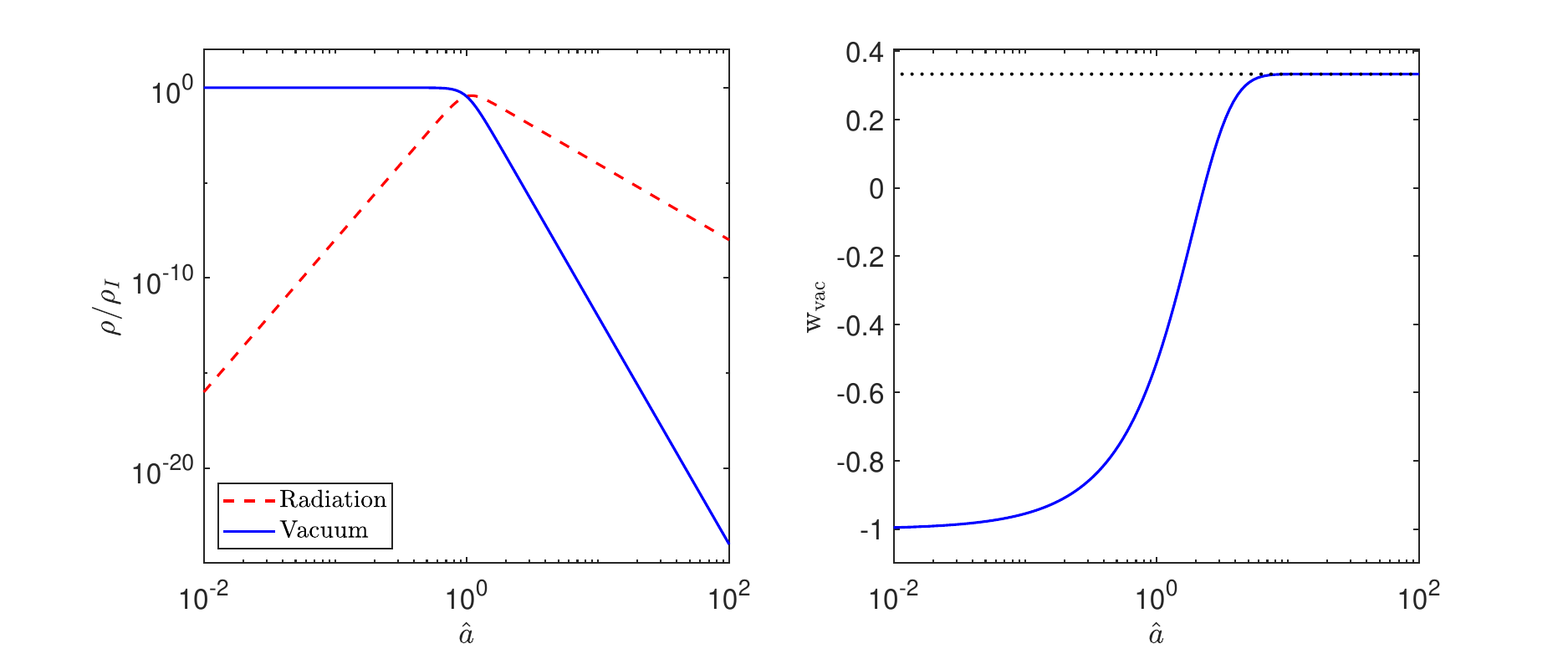}
\end{center}
\caption{RVM-inflation. On the left, the vacuum and relativistic matter densities \eqref{eq:Inflationdensities} before and after the transition point `vacuum$\to$radiation': $\hat{a}\equiv a/a_\star=1$ ($a_*\sim 10^{-29}$).  The VED ($\simeq$const.) decays fast into radiation.  On the right, the vacuum EoS transits from $w_{\rm vac}\simeq -1$ into an incipient radiation phase and adopts its EoS: $w_{\rm vac}\rightarrow 1/3$; hereafter the standard FLRW regime starts.}
\end{figure}
\end{center}
We have defined $\hat{a}\equiv a/a_*$, where $a_*$ determines the transition point into radiation-dominated epoch from a vacuum dominated epoch. It is estimated to be around $a_*\sim 10^{-29}-10^{-30}$, see \cite{Yu2020}. At the beginning of inflation, the Hubble rate evolves very little and remains close to $H_{\rm I}\sim\sqrt{M_{Pl}M_X}$. Similarly, $\rho_{\rm vac}$ stays close to $\rho_{\rm I} \sim M_{\rm Pl}^3 M_X$, whereas the radiation energy is negligible during inflation since  $\hat{a}^{8}\lll$ for $\hat{a}\ll1$. In contrast, we do recover the usual behavior of radiation during the current FLRW era. Indeed, for  $ \hat{a} \gg 1$ Eq.\,\eqref{eq:Inflationdensities} implies $\rho_r\sim a^{-4}$. However, to obtain $\rho_{\rm vac}\to\rho_{\rm vac}^0$ for the current universe using these inflationary formulas is harder. It would require to keep an additive constant in the low energy regime, so as to analytically interpolate with the high energy regime. Solving the equations is then more difficult, but it is actually unnecessary since the low energy solution was already found, Eq.\,\eqref{eq:RVMLate}, and the term is there!  The higher powers in the adiabatic expansion rapidly decay and become irrelevant; and, as said, without disturbing the primordial Big Bang Nucleosynthesis physics\cite{Asimakis2022}. A graphical description of RVM-inflation where the above features are implemented can be appraised in Figure 1.

During inflation, the equation of state (EoS) of the quantum vacuum, $w_{\rm vac}=P_{\rm vac}/\rho_{\rm vac}$, remains essentially -1 (cf. Fig. 1, right plot). The evaluation of the EoS requires the explicit computation of the vacuum pressure as well, $P_{\rm vac}$,  which differs from the VED by terms which  depend on the derivatives of $H$, rather than on $H$ itself, and hence vanish for $H=$const. This explains the mentioned result\cite{CristianJoan2022a}. After inflation, $H$ is no longer constant  and a dynamical evolution of vacuum's EoS is expected, which is considered in the next section.

\section{Dynamical equation of state for the running vacuum}\label{sect:Eos}

As indicated previously, we should not presume that the EoS of the quantum vacuum is exactly $P_{\rm vac} = -\rho_{\rm vac}$ in QFT  since this formula receives quantum corrections. Obviously the vacuum EoS cannot depart too much from the traditional value $-1$, but the small deviation opens up the possibility of observing effective quintessence or phantom behavior from the vacuum itself!  The vacuum pressure is defined in a way similar to the vacuum energy density. Assuming the vacuum to be an homogeneous and isotropic medium (it should preserve the Cosmological Principle) we may define the pressure using any diagonal $ii$-component of the renormalized vacuum stress tensor:
\begin{equation}
P_{\rm vac}=-\rho_{\Lambda}+\frac{\langle T_{11}^{\delta\phi}\rangle_{\rm ren}}{a^2}\,.
\end{equation}
Following an analogous computation to the one of the vacuum energy density we can derive the quantum effects on the vacuum EoS (see\,\cite{CristianJoan2022a,CristianJoan2022b}  for more details). A sufficiently accurate expression in terms of the redshift can be cast as follows:
\begin{equation}\label{eq:w_vac}
 w_{\rm vac}(z) = -1 +\frac{\nu_{\rm eff}\left(\Omega_{\rm m}^0 (1+z)^3+\frac{4}{3}\Omega_{\rm r}^0 (1+z)^4\right)}{\Omega_{\rm vac}^0+\nu_{\rm eff}\left(-1+\Omega_{\rm m}^0 (1+z)^3 +\Omega_{\rm r}(1+z)^4+\Omega_{\rm vac}^0 \right)}\,,
\end{equation}
being $\Omega_{\rm m}^0 \sim 0.3$, $\Omega_{\Lambda}^0 \sim 0.7$ and $\Omega_{\rm r}^0 \sim 10^{-4}$  the current energy fractions of matter, vacuum fluid and radiation. The deviations from $-1$ in the EoS formula are proportional to the same small parameter $\nu_{\rm eff}$ which is involved in the running of the VED. The value of $\nu_{\rm eff}$, as we know,  has been  fitted to the cosmological data\,\cite{RVMPheno1}.  Recalling that $z_{\rm eq}= \frac{\Omega_{\rm m}^0}{\Omega_{\rm r}^0}-1\approx 3300$ determines the equality point between matter and radiation, the vacuum equation of state \eqref{eq:w_vac} can be displayed in three regimes:
\begin{equation}
w_{\rm vac} (z)= \left\{ \begin{array}{lcc} \frac{1}{3} & \textrm{for} & z \gg z_{\rm eq}\textrm{ and }\Omega_{\rm r}^0 (1+z) \gg \Omega_{\rm m}^0 , \\  0 & \textrm{for} & \mathcal{O}(1)<z < z_{\rm eq}\textrm{ and }\Omega_{\rm r}^0 (1+z) \ll \Omega_{\rm m}^0  , \\  -1+\nu_{\rm eff}\frac{\Omega_{\rm m}^0}{\Omega_{\rm vac}^0}(1+z)^3 & &  \textrm{for} -1<z < \mathcal{O}(1). \end{array} \right.
\end{equation}
\begin{center}
\begin{figure}[t]\label{figure:EoSvac}
\begin{center}
\includegraphics[scale=0.6]{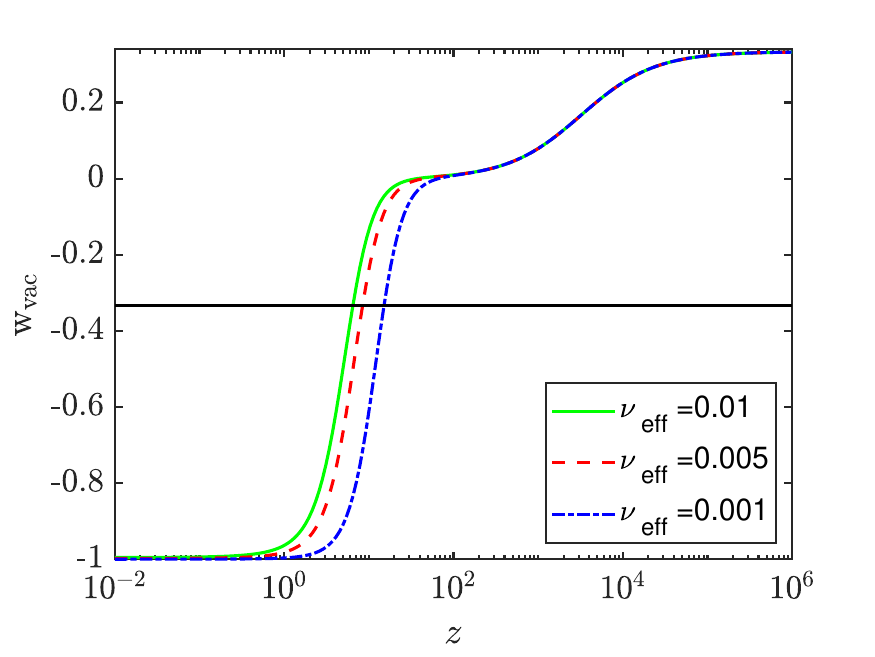}
\end{center}
\caption{The equation of state of the (quantum) vacuum as a function of the redshift,  from the radiation- and matter-dominated epochs up to the present time, for different values of $\nu_{\rm eff}$.}
\end{figure}
\end{center}
The quantum vacuum, therefore,  behaves as a cosmic chameleon since it imitates the dominant energy component at each cosmic epoch: in fact, its EoS is $-1$ during inflation (cf. Fig. 1),  1/3 during radiation dominated epoch, remains near 0 during matter dominated epoch and, finally,  approaches closely (but not exaclty) -1 during the current era, the difference being of order $\nu_{\rm eff}$. Depending on the sign of $\nu_{\rm eff}$, the vacuum can either behave as quintessence ($\nu_{\rm eff}>0$) or phantom ($\nu_{\rm eff}<0$). The three distinct  (post-inflationary) regimes can be clearly seen in Figure 2.

Finally, we remark that the combined running of the VED, $\rho_{\rm vac}(H)$, and of the gravitational coupling, $G_N(H)$, can have a synergic impact on the $H_0$ tension. The simultaneous running of the VED and of $G_N$ is a consequence of the Bianchi identity, which links them both.  The running of $G_N(H)$ with $H$ is very mild (in fact, logarithmic) but it is collaborative. At the end of the day, we can assert that the quantum properties of the vacuum can have a non-negligible positive bearing on the overall fit to the cosmological data, including a significant alleviation of the $\sigma_8$ and $H_0$ tensions --  see \cite{RVMPheno1,RVMPheno2,BDRVM} for the phenomenological analyses.

\section{Discussion and conclusions}\label{sec:conclusions}

In this work, we have presented a consistent discussion of the vacuum energy in quantum field theory (QFT) in  FLRW spacetime, together with the corresponding implications on the cosmological constant problem (CCP).  The new approach constitutes the running vacuum model (RVM) approach to cosmology\,\cite{JSPRev2022}, which puts on firm QFT ground a panoply of ideas collected over the years on the possibility that the cosmic vacuum can be dynamical, see \cite{JSPRev2013,JSPRev2015}  and references therein.  For more details on the technical QFT part, see \cite{CristianJoan2020,CristianJoan2022a,CristianJoan2022b,CristianJoanSamira2023}.  From the calculation of the renormalized energy-momentum tensor (EMT) of a quantized scalar field non-minimally coupled to the FLRW background\cite{CristianJoan2020} (which can also be extended to fermions\cite{CristianJoanSamira2023}) we have identified new important properties of the quantum vacuum which can alleviate the  CCP and also the pressing issues on the cosmological tensions that have been gripping us for a long time.

The RVM approach is based on adiabatic regularization and  renormalization of the EMT, starting from the WKB expansion of the field modes in curved spacetime. The renormalized EMT is then defined as the difference between its on-shell value and its value at an arbitrary renormalization point $M$ up to 4{\it th} adiabatic order. The value $M$ plays the role of floating scale of the RVM renormalizatuion framework. The different epochs of the cosmic history (characterized by the value of the Hubble rate, $H$) are explored through the physical setting $M=H$. The renormalized vacuum energy density (VED) is the sum of the renormalized EMT and the renormalized $\rL$ term in the Einstein-Hilbert action. The VED appears as an expansion in  (even) powers of the Hubble function and its derivatives, $\rho_{\rm vac}(H, \dot{H},...)$. It adopts the canonical form foreseen from long existing notions on the RVM (cf. \cite{JSPRev2013} and references therein). We have also found extended QFT implementations\cite{CristianJoan2022a} of the RVM where Newton's `constant' is also running with the expansion, $G_N(H)$. Overall we have proven, and certainly improved, the traditional RVM form of the VED. The old ideas have been around for many years, sustained by semi-qualitative renormalization group arguments, but only after the QFT work of\cite{CristianJoan2020,CristianJoan2022a,CristianJoan2022b,CristianJoanSamira2023} the RVM has finally been firmly put on a fundamental QFT ground.

The basic message is that neither  the VED at the present time, $\rho_{\rm vac}^0 =\rho_{\rm }(H_0)$, nor the associated cosmological constant, $\Lambda=8\pi \rho_{\rm vac}(H_0) G(H_0)$, are really constants of nature, since $H_0$ is to be replaced by a generic $H$. Hence they are both dynamical quantities that evolve with the cosmic expansion. But the dynamics is smooth enough and  hence the behavior is still $\Lambda$CDM-like, at least in the late Universe.  The evolution of the VED between two nearby epochs in the recent history is $\delta \rho_{\rho_{\rm} vac}\propto \nu_{\rm eff}H^2$. The small parameter  $|\nu_{\rm eff}| \ll 1$ is QFT-calculable. It plays the role of the $\beta$-function of the runnign VED.  Remarkably, the latter is free from the undesired  $\sim m^4$ contributions, which if present would awfully recreate the need for extreme fine tuning associated with the CCP.  Obviously, this property is a very important theoretical advance of the RVM approach.

No less important are the consequences on the phenomenological/observational side. In  previous works, the RVM has been successfully confronted against a large number of cosmological data. The effective parameter $\nu_{\rm eff}$ has been fitted to values around $\sim 10^{-3}$, see \cite{RVMPheno1}  and \cite{RVMPheno2}. The same analyses show that the $H_0$ and growth tensions can be alleviated\cite{RVMPheno1}. Besides, RVM extensions into Brans-Dicke type theories prove also fruitful to cut down the cosmological tensions\,\cite{BDRVM}.

In another vein, the higher order terms $H^4, H^6,...$ may play a major role in the very early universe. They can naturally enhance the magnitude of the VED and provide a new mechanism for inflation, called `RVM-inflation',\cite{CristianJoan2022a,CristianJoan2022b}, which consistently describes the transition of the high energy energy densities and entropy\cite{Sola:2015csa,Yu2020} of the very early universe into the FLRW regime, whereby implementing the graceful exit mechanism into the radiation-dominated epoch (RDE) up to our days.

Another remarkable prediction of the QFT calculations in the RVM renormalziation framework is that the vacuum  EoS is not equal to -1 throughout the cosmic history\,\cite{CristianJoan2022b}. It changes with the redshift: it behaves as a true cosmological constant during inflation ($w_{\rm vac}^{\inf} =-1$), mimicking radiation during the RDE ($w_{\rm vac} \approx 1/3$), behaving as dust ($w_{\rm vac}\approx 0$) during the matter dominated-epoch; and, no less remarkable, effectively behaving as quintessence ($w_{\rm vac}(z)\gtrsim  -1$) or phantom DE ($w_{\rm vac}(z)\lesssim-1$) around our time, depending on the sign of $\nu_{\rm eff}$.  This feature shows that the vacuum dynamics could be the ultimate responsible for the long sought-after dynamical DE, vigorously revived  anew from the recent DESI measurements\cite{DESI:2024mwx}. Therefore, ad hoc scalars fields for describing the DE could just be dispensable. We remark that our conclusions have been obtained upon rigorous computation within the framework of QFT in curved spacetime.

\vspace{-0.1cm}

\section*{Acknowledgements}
This work is  partially supported by grants PID2022-136224NB-C21 and  PID2019-105614GB-C21, from MCIN/AEI/10.13039/501100011033.  JSP is funded also  by  2021-SGR-249 (Generalitat de Catalunya) and
CEX2019-000918-M (ICCUB, Barcelona). We acknowledge networking support by the COST Association Action CA21136 ``{\it Addressing observational tensions in cosmology
with systematics and fundamental physics (CosmoVerse)}''.

\bibliographystyle{ws-procs961x669}

\end{document}